%% file: recoil_he_rom_v6_sub.tex
\documentclass[preprint,aps,prl,showpacs,superscriptaddress,floatfix]{revtex4}
\usepackage{graphicx}
\usepackage{times}
\usepackage{nicefrac}
\usepackage{amsmath}
\usepackage{amsfonts}
\usepackage{amssymb}
\usepackage{amsthm}
\usepackage{epsf}
\usepackage{bm}
\usepackage{times}

\usepackage{booktabs}
\usepackage{longtable}
\usepackage{siunitx}

\newcommand{\bfr}{{\bf r}}
\newcommand{\bfD}{{\bf D}}
\newcommand{\bfx}{{\bf x}}

\newcommand{\bfp}{{\bf p}}

\newcommand{\balpha}{{\mbox{\boldmath$\alpha$}}}
\newcommand{\bnabla}{{\mbox{\boldmath$\nabla$}}}

\newcommand{\be}{\begin{eqnarray}}
\newcommand{\ee}{\end{eqnarray}}
\newcommand{\la}{\langle}
\newcommand{\ra}{\rangle}

\newcommand{\veps}{\varepsilon}

\newcommand{\matr}[3]{\langle #1 | #2 | #3 \rangle}        

\usepackage{ulem}
\usepackage{color}

\begin{document}

\title{Nuclear recoil effect on the binding energies in  highly charged He-like ions}

\author{A. V. Malyshev}

\affiliation {Department of Physics, St.~Petersburg State University,
Universitetskaya 7/9,
199034 St.~Petersburg, Russia}

\author{R. V. Popov}

\affiliation {Department of Physics, St.~Petersburg State University,
Universitetskaya 7/9,
199034 St.~Petersburg, Russia}

\author{V. M. Shabaev}

\affiliation {Department of Physics, St.~Petersburg State University,
Universitetskaya 7/9,
199034 St.~Petersburg, Russia}

\author{N. A. Zubova}

\affiliation {Department of Physics, St.~Petersburg State University,
Universitetskaya 7/9,
199034 St.~Petersburg, Russia}

\begin{abstract}

The most precise to-date evaluation of the nuclear recoil effect on the $n=1$ and $n=2$ energy levels of He-like ions is presented in the range $Z=12-100$. The one-electron recoil contribution is calculated  within the framework of the rigorous QED approach to first order in the electron-to-nucleus mass ratio~$m/M$ and to all orders in the parameter~$\alpha Z$. The two-electron $m/M$~recoil term is calculated employing the $1/Z$~perturbation theory. The  recoil contribution of the zeroth order in $1/Z$ is evaluated to all orders in~$\alpha Z$, while the $1/Z$~term is calculated using the Breit approximation. The recoil corrections of the second and higher orders in~$1/Z$ are taken into account within the nonrelativistic approach. The obtained results are compared with the previous evaluation of this effect [A.~N.~Artemyev \textit{et al.}, Phys. Rev. A {\bf 71}, 062104 (2005)].

\end{abstract}
\pacs{31.30.J-, 12.20.Ds}
\maketitle

\section{Introduction}

Highly charged ions provide a unique opportunity to probe the methods of quantum electrodynamics (QED) in presence of strong electromagnetic fields induced by heavy nuclei. To date, the most stringent tests of the QED effects on the binding energies of heavy few-electron ions are associated with the Lamb shift measurements in hydrogenlike and lithiumlike uranium ions~\cite{Stoehlker:2000:3109,Gumberidze:2005:223001,Schweppe:1991:1434,Brandau:2003:073202,Beiersdorfer:2005:233003} (for the related theory, see Refs.~\cite{Yerokhin:2006:253004,Kozhedub:2008:032501} and references therein). However, these results alone are not sufficient for the full-fledged comparison between theory and experiment, especially in view of a systematic discrepancy between the theoretical predictions and experimental data that has been claimed recently for middle-$Z$ heliumlike ions~\cite{Chantler:2012:153001}. To date, the most accurate calculation of the $n=1$ and $n=2$ energies in heliumlike ions was performed in Ref.~\cite{Artemyev:2005:062104}. Compared to the previous evaluations of the binding energies in He-like ions~\cite{Drake:1988:586,Johnson:1992:R2197,Chen:1993:3692,Plante:1994:3519,Indelicato:1995:1132}, the calculation by Artemyev {\it et al.}~\cite{Artemyev:2005:062104} included all two-electron Feynman diagrams up to the second order of the QED perturbation theory in the fine structure constant $\alpha$ without an expansion in the parameter $\alpha Z$. The theoretical predictions of this work are in agreement with a number of the experimental results (see, e.g., corresponding references and comparison of theory and experiment in Refs.~\cite{Artemyev:2005:062104,Trassinelli:2007:129,Trassinelli:2009:63001,Amaro:2012:043005,Rudolph:2013:103002,Kubicek:2014:032508,Beiersdorfer:2015:032514,Epp:2015:020502_R,Holmberg:2015:012509,Machado:HeBe:preprint}). However, on the grounds of new sensitive measurements with heliumlike titanium ($Z=22$) and a statistical analysis of all the data available in literature, Chantler \textit{et al.}~\cite{Chantler:2012:153001} have claimed that there exists a systematic discrepancy between theory and experiment, which behaves approximately as~$Z^3$. Therefore, in addition to new  measurements of the transition energies in He-like ions, further improvements of the theoretical predictions are very desirable. With this paper, we start systematic revision and improvement of the calculations presented in Ref.~\cite{Artemyev:2005:062104}. Namely, we perform the most accurate evaluation of the nuclear recoil corrections to the $n=1$ and $n=2$ energy levels  in He-like ions in the range $Z=12-100$.  

Relativistic units ($\hbar=c=1$) are used in the paper.

\section{Basic formulas}

In this paper the calculations of the nuclear recoil corrections to the binding energies of highly charged He-like ions are performed using perturbation theory in the parameter~$1/Z$. In the $jj$~coupling, the unperturbed wave function can be written as
\be \label{manydet}
u_i(\bfx_1,\bfx_2)=
A_N \sum_{m_{i_1} m_{i_2}}\langle \, j_{i_1} m_{i_1} j_{i_2} m_{i_2}\,|\,JM_J \, \rangle
\sum_{P}(-1)^P \psi_{Pi_1}(\bfx_1)\psi_{Pi_2}(\bfx_2)\,,
\ee
where $ \psi_{k}(\bfx) $ are one-electron Dirac wave functions, $P$ is the permutation operator, $(-1)^{P}$ is the permutation parity, $A_N$ is the normalization factor equal to $1/\sqrt{2}$ for non-equivalent electrons and to $1/2$ for equivalent electrons, $J$ is the total angular momentum, and $M_J$ is its projection. In the simplest case of a one-determinant wave function, we have
\begin{equation}
u_{i}(\bfx_1,\bfx_2)
=\frac{1}{\sqrt{2}}\sum_{P}(-1)^{P}\psi_{Pi_1}({\bf x_{1}})
        \psi_{Pi_2 }({\bf x_{2}})\,.
    \label{e3e1}
\end{equation}  

The fully relativistic theory of the nuclear recoil effect in atoms can be formulated only within the framework of quantum electrodynamics~\cite{Shabaev:1985:394:note,Shabaev:1988:107:note,Pachucki:1995:1854,Shabaev:1998:59,Adkins:2007:042508}. The recoil effect on the binding energy of a He-like ion is given by a sum of one-electron and two-electron parts. The one-electron part is obtained by summing  the one-electron recoil contributions for both electrons. These contributions to first order in the electron-to-nucleus mass ratio~$m/M$, to zeroth order in~$\alpha$, and to all orders in~$\alpha Z$, are conveniently represented by a sum of low-order and  higher-order terms, $\Delta E=\Delta E_{\rm L}+\Delta E_{\rm H}$, where~\cite{Shabaev:1985:394:note}
\be \label{low}
\Delta E_{\rm L}&=&
\frac{1}{2M} \la a|
\left[ \bfp^2
 -\frac{\alpha Z}{r}\left(\balpha+\frac{(\balpha\cdot\bfr)\bfr}{r^2}\right)
  \cdot\bfp
  \right]|a\ra 
\ee
\be  \label{high}
\Delta E_{\rm H}&=&\frac{1}{M}
\frac{i}{2\pi} \int_{-\infty}^{\infty} \! d\omega\, \la
 a|\left(\bfD(\omega)-\frac{[\bfp,V]}{\omega+i0}\right)
 G(\omega+\veps_a)\left(\bfD(\omega)+\frac{[\bfp,V]}{\omega+i0}\right)|a\ra\,.
\ee
Here $\bfp=-i\bnabla$ is the momentum operator, $V(r)=-\alpha Z/r$ is the Coulomb potential of the nucleus, $|a\ra$ is the Dirac wave function $\psi_{a}(\bfx)$ for the potential $V(r)$, $\veps_a$ is the corresponding Dirac energy, $G(\omega)=\sum_n|n\ra \la n|[\omega-\veps_n(1-i0)]^{-1}$ is the Dirac-Coulomb Green function, $D^k(\omega)=-4\pi\alpha Z\alpha^l D^{lk}(\omega)$,
\begin{eqnarray} \label{06photon}
D^{lk}(\omega,{\bf r})=-\frac{1}{4\pi}\left\{\frac
{\exp{(i|\omega|r)}}{r}\delta_{lk}+\nabla^{l}\nabla^{k}
\frac{(\exp{(i|\omega|r)}
-1)}{\omega^{2}r}\right\}\,
\end{eqnarray}
is the transverse part of the photon propagator in the Coulomb gauge,  $\balpha$ is a vector of the Dirac matrices, and the summation over the repeated indices is implied. In Eq.~(\ref{high}) and below the scalar product is implicit. The low-order term can be derived from the relativistic Breit equation, while the derivation of the higher-order term requires using QED beyond the Breit approximation. For this reason, we refer to them as the non-QED and QED one-electron contributions, respectively. 

For the case of a one-determinant wave function constructed from the Dirac bispinors $\psi_{a}(\bfx)$ and $\psi_{b}(\bfx)$, the two-electron recoil contribution to zeroth order in $1/Z$ is given by~\cite{Shabaev:1988:107:note}
\begin{eqnarray} \label{two}
\Delta E^{(\rm 2el)}=
\frac{1}{M} \sum_{P}(-1)^{P}
\langle Pa|[\bfp-\bfD
(\varepsilon_{Pa}-\varepsilon_{a})]
|a\rangle
\langle Pb|[\bfp-\bfD
(\varepsilon_{Pb}-\varepsilon_{b})]
|b\rangle\,.
\end{eqnarray}
The transition to the general case of a many-determinant function~(\ref{manydet}) causes no problem and can be done in the final expression for the energy shift. For quasidegenerate $(1s2p_{1/2})_1$ and $(1s2p_{3/2})_1$ states one has to construct a matrix~$H$, which plays the role of the  Shr\"odinger-like Hamiltonian acting in the space of the corresponding unperturbed states~\cite{TTGF}. The matrix elements of the recoil contribution between the one-determinant wave functions $u_i$ and $u_k$ with the unperturbed energies $E_i^{(0)}=\veps_{i_1} + \veps_{i_2}$ and $E_k^{(0)}=\veps_{k_1} + \veps_{k_2}$ read as
\be \label{eq:2el_rec_quasi}
H_{ik}^{\rm 2el} &\!\!=\!\!&  \frac{1}{2M} \, \sum_P (-1)^P
   \Bigg\{ \matr{Pi_1}{\,\big[\bfp-\bfD(\Delta_1)\big]\,}{k_1} 
           \matr{Pi_2}{\,\big[\bfp-\bfD(\Delta_1)\big]\,}{k_2}                \nonumber \\
&& \qquad\qquad\qquad  
     \,  + \matr{Pi_1}{\,\big[\bfp-\bfD(\Delta_2)\big]\,}{k_1} 
           \matr{Pi_2}{\,\big[\bfp-\bfD(\Delta_2)\big]\,}{k_2} \Bigg\}  \, ,
\ee
where $\Delta_1 = \veps_{Pi_1} - \veps_{k_1}$ and $\Delta_2 = \veps_{Pi_2} - \veps_{k_2}$. The summation over the angular momentum projections with the Clebsch-Gordan coefficients in Eq.~(\ref{eq:2el_rec_quasi}) in order to obtain the states with the given total angular moment $J$ is also straightforward. The derivation of the formula~(\ref{eq:2el_rec_quasi}) is similar to the derivation of the expression for the one-photon exchange contribution in the case of quasidegenerate levels, see, e.g., Refs.~\cite{Artemyev:2005:062104,TTGF}.

To evaluate the two-electron contributions of the first and higher orders in~$1/Z$, one can use the effective two-electron recoil operator,
\be \label{br1}
H_M=\frac{1}{2M}\sum_{i,k}\left[\bfp_i\cdot \bfp_k
  -\frac{\alpha Z}{r_i}\left(\balpha_i+\frac{(\balpha_i\cdot\bfr_i)\bfr_i}{r_i^2}\right)
  \cdot\bfp_k\right]\,,
\ee
which is obtained as a sum of the low-order one-electron operator~(\ref{low}) and the zero-energy-transfer limit of the expression~(\ref{two}), omitting the term with two $\bfD$~operators, which contributes to the higher orders. The operator~(\ref{br1}) can be also derived by reformulating Stone's theory~\cite{Palmer:1987:5987}. Within the lowest-order relativistic (Breit) approximation the recoil effect on the binding energy to the first order in~$m/M$ can be evaluated by averaging the operator~(\ref{br1}) with the wave functions obtained from the Dirac-Coulomb-Breit Hamiltonian. Nowadays, the Hamiltonian $H_M$ is widely used in relativistic calculations of the isotope shifts (see, e.g., Refs.~\cite{Shabaev:1994:1307,Tupitsyn:2003:022511,Orts:2006:103002,Korol:2007:022103,Gaidamauskas:2011:175003,Naze:2013:2187,Zubova:2016:052502,Filippin:2017:042502} and references therein). In the next section, it will be used to evaluate the  $1/Z$ contribution to the recoil effect in the framework of the Breit approximation.

\section{Numerical calculations}

For the point nucleus case, the low-order one-electron term $\Delta E_{\rm L}$ can be evaluated analytically~\cite{Shabaev:1985:394:note}:
\begin{eqnarray} \label{06shabaeveq112}
\Delta  E_{\rm L}^{\rm(p.n.)}=\frac{m^2-\veps^2}{2M}\,,
\end{eqnarray}
where $\veps$ is the Dirac energy. The higher-order one-electron (QED) contribution $\Delta E_{\rm H}$ was evaluated numerically for point nuclei  in Refs.~\cite{Artemyev:1995:1884,Artemyev:1995:5201,Adkins:2007:042508}. The formulas~(\ref{low}) and (\ref{high}) can also be employed for  calculations which partially account for the nuclear size correction to the recoil effect. This can be done by replacing the pure Coulomb potential  of the nucleus $V=-\alpha Z/r$ with the potential of an extended nucleus. These calculations  were carried out for  $1s$ and $2s$ states in Refs.~\cite{Shabaev:1998:4235,Shabaev:1999:493}. In the present paper, we have performed the corresponding calculations of the low- and higher-order terms for one-electron $1s$, $2s$, $2p_{1/2}$, and  $2p_{3/2}$ states using the Fermi model for the nuclear charge distribution. The nuclear charge radii were taken from Refs.~\cite{Angeli:2013:69,Yerokhin:2015:033103}. The summations over intermediate electron states have been performed using the dual-kinetic-balance finite basis set method~\cite{splines:DKB} with the basis functions constructed from B-splines~\cite{Sapirstein:1996:5213}. The results of the calculations, expressed in terms of the function $A(\alpha Z)$,
\begin{eqnarray}\label{A}
\Delta E = (\alpha Z)^2\frac{m^2}{M}A(\alpha Z)\,,
\end{eqnarray}
are presented in  Table~\ref{H-like}. For  $1s$ and $2s$ states, the obtained results  agree with the previous calculations~\cite{Shabaev:1998:4235,Shabaev:1999:493} but have a higher accuracy for heavy ions, especially in case of $2s$ state. The rigorous treatment of the finite nuclear size correction to the recoil effect is presently accessible only within the framework of the Breit approximation~\cite{Grotch:1969:350,Borie:1982:67,Aleksandrov:2015:144004}. The correction  $\delta A_{\rm L}^{\rm (NS,add)} $, which determines the difference between the exact treatment of the nuclear size correction to the low-order recoil effect and its evaluation by formula~(\ref{low}) with the extended nucleus potential~\cite{Aleksandrov:2015:144004}, is also shown in Table~\ref{H-like}. The uncertainty of the total one-electron contribution, presented in the table, is due to the approximate treatment of the nuclear size effect on the QED recoil contribution. It is estimated as a product of the relative value of the corresponding correction within the Breit approximation and the QED term.

\begin{figure}
\begin{center}
\includegraphics[width=13cm]{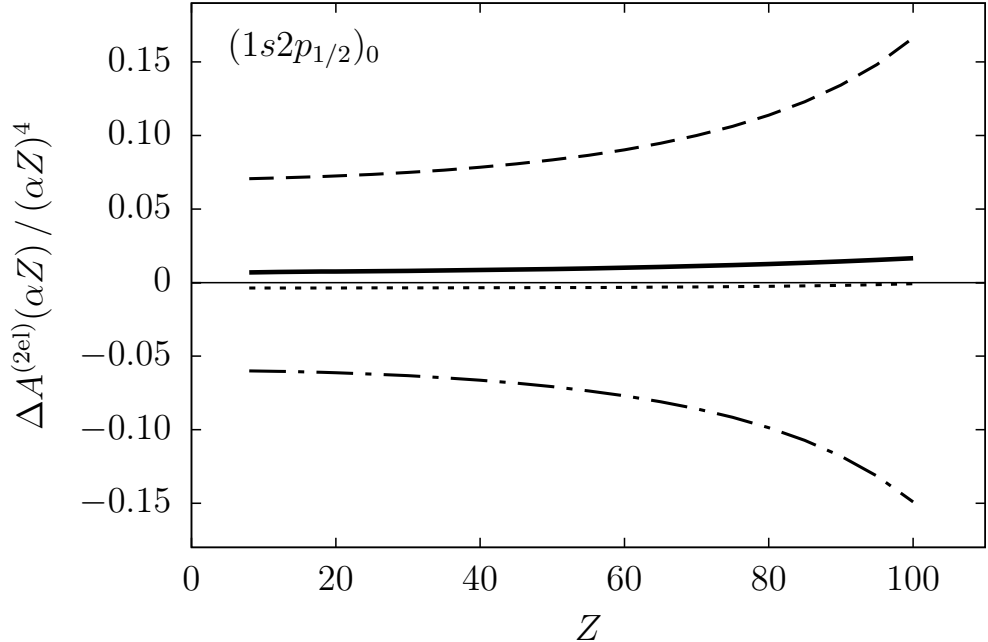}
\caption{\label{fig:1s2p1_0}
The two-electron $m/M$ recoil contribution of sixth and higher orders in $\alpha Z$ \, ($\sim(\alpha Z)^{6+}m^2/M$) for $(1s 2p_{1/2})_0$ state. The dashed line represents the higher-order correction (in $\alpha Z$) to the main term corresponding to the operator~(\ref{br1}). The dotted line shows the frequency-dependent correction to the term with one $\bfD$ operator in Eq.~(\ref{eq:2el_rec_quasi}). The dash-dotted line indicates the contribution of the term with two operators $\bfD$ in Eq.~(\ref{eq:2el_rec_quasi}). Finally, the solid line stands for the total higher-order correction to the two-electron recoil contribution. All the contributions are presented in terms of the function $\Delta A^{(\rm 2el)}(\alpha Z)$, see Eq.~(\ref{eq:A_expan}), divided by $(\alpha Z)^4$.}
\end{center}
\end{figure}

\begin{figure}
\begin{center}
\includegraphics[width=13cm]{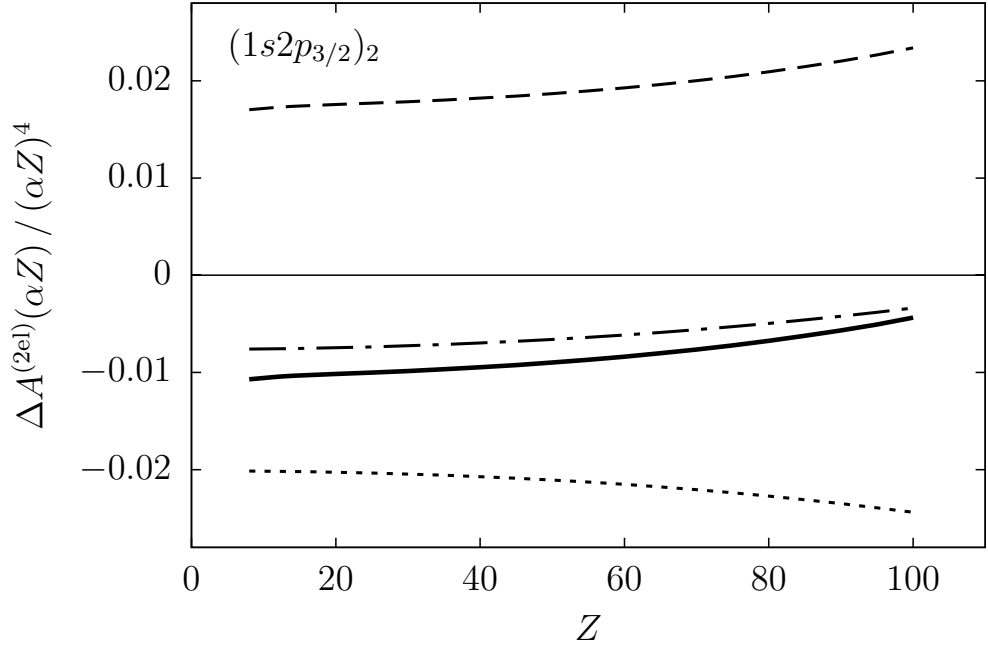}
\caption{\label{fig:1s2p3_2}
The two-electron $m/M$ recoil contribution of sixth and higher orders in $\alpha Z$ \, ($\sim(\alpha Z)^{6+}m^2/M$) for $(1s 2p_{3/2})_2$ state. Notations are the same as in Fig.~\ref{fig:1s2p1_0}.}
\end{center}
\end{figure}

\begin{figure}
\begin{center}
\includegraphics[width=13cm]{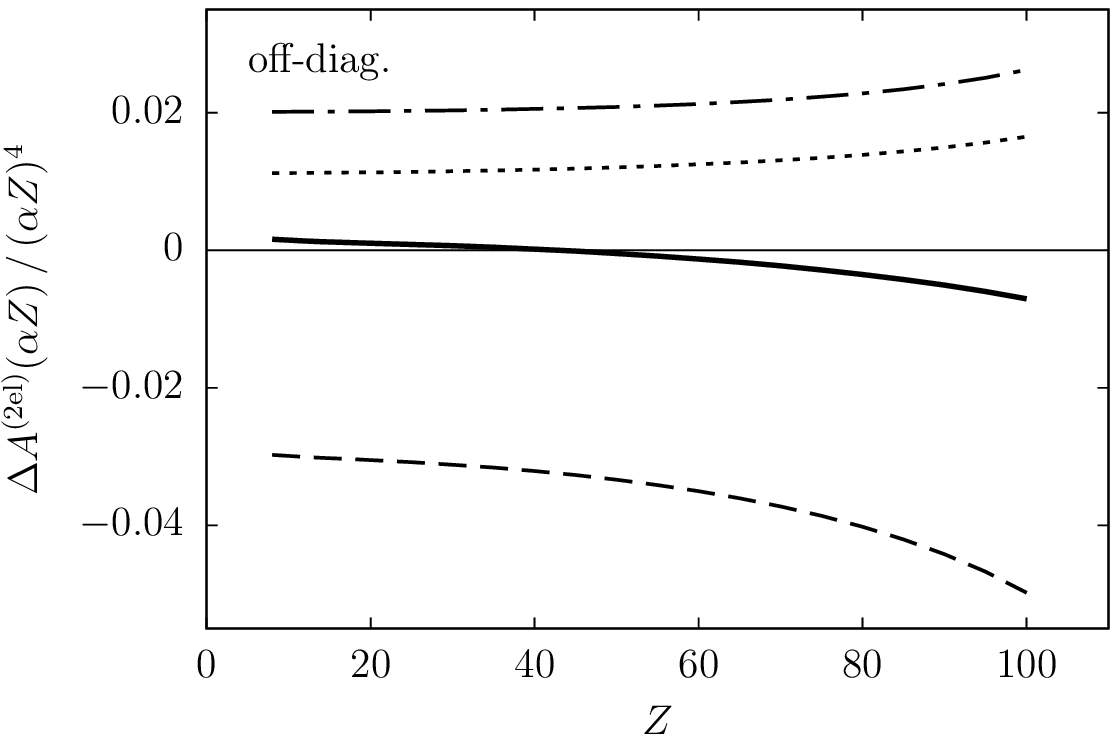}
\caption{\label{fig:fig:quasi_12}
The two-electron $m/M$ recoil contribution of sixth and higher orders in $\alpha Z$ \, ($\sim(\alpha Z)^{6+}m^2/M$) for the off-diagonal matrix element. Notations are the same as in Fig.~\ref{fig:1s2p1_0}.}
\end{center}
\end{figure}

To zeroth order in $1/Z$, the two-electron recoil contribution is  evaluated to all orders in $\alpha Z$ according to formula~(\ref{two}) and its generalization for quasidegenerate states~(\ref{eq:2el_rec_quasi}). Performing the summation over the angular momentum projections in Eqs.~(\ref{two}) and (\ref{eq:2el_rec_quasi}), one can derive the following relation between the two-electron recoil contributions for $(1s 2p_{1/2})_0$ and $(1s 2p_{1/2})_1$ states in He-like and $(1s^2)2p_{1/2}$ state in Li-like ions 
\be \label{eq:dE_p1_2el_relation}
 \Delta E^{(\rm 2el)}_{(1s^2)2p_{1/2}} =
 \Delta E^{(\rm 2el)}_{(1s 2p_{1/2})_0} = 
3\Delta E^{(\rm 2el)}_{(1s 2p_{1/2})_1} \, .
\ee
The analogous relation takes place for the levels involving $2p_{3/2}$ electron,
\be \label{eq:dE_p3_2el_relation}
  \Delta E^{(\rm 2el)}_{(1s^2)2p_{3/2}} =
  \Delta E^{(\rm 2el)}_{(1s 2p_{3/2})_2} = 
-3\Delta E^{(\rm 2el)}_{(1s 2p_{3/2})_1} \, .
\ee
In Eqs.~(\ref{eq:dE_p1_2el_relation}) and (\ref{eq:dE_p3_2el_relation}), $\Delta E^{(\rm 2el)}_{(1s 2p_{1/2})_1}$ and $\Delta E^{(\rm 2el)}_{(1s 2p_{3/2})_1}$ denote the corresponding diagonal matrix elements of the operator~(\ref{eq:2el_rec_quasi}). The two-electron recoil contribution to zeroth order in $1/Z$ can be divided into two parts: the main part which corresponds to the zero-energy-transfer limit, given by the operator~(\ref{br1}), and the frequency-dependent correction, which also incorporates the term with two operators $\bfD$ in Eqs.~(\ref{two}) and (\ref{eq:2el_rec_quasi}). The related contributions, expressed in terms of the function $A(\alpha Z)$ defined by equation~(\ref{A}), are presented in the second and third columns of Table~\ref{two-electron_0}. For the two-electron recoil contribution the $\alpha Z$ expansion for the function $A(\alpha Z)$ reads as
\be\label{eq:A_expan}
   A^{(\rm 2el)}(\alpha Z) = a_0 + a_2 (\alpha Z)^2 + \Delta A^{(\rm 2el)}(\alpha Z) \, .
\ee
For comparison, in the last column of Table~\ref{two-electron_0} we list the $\alpha Z$-expansion results, $a_0 + a_2 (\alpha Z)^2$, obtained analytically for the point nucleus case in Ref.~\cite{Shabaev:1994:1307}. It is surprising that, while the frequency-dependent part, which contributes beyond this order, is very significant for heavy ions, the complete $\alpha Z$-dependent results $A^{(\rm 2el)}(\alpha Z)$ are rather close to the $\alpha Z$-expansion results $a_0 + a_2 (\alpha Z)^2$, used in Ref.~\cite{Artemyev:2005:062104}. In order to illustrate this fact, we present the corrections $\Delta A^{(\rm 2el)}(\alpha Z)$ to the two-electron recoil contributions for $(1s 2p_{1/2})_0$ and $(1s 2p_{3/2})_2$ states and for the off-diagonal matrix element between $(1s 2p_{1/2})_1$ and $(1s 2p_{3/2})_1$ states in Figs.~\ref{fig:1s2p1_0}, \ref{fig:1s2p3_2}, and \ref{fig:fig:quasi_12}, respectively. The total higher-order corrections, which are shown with  solid lines, are obtained by summing the higher-order Breit corrections resulting from the calculations with the operator~(\ref{br1}) (dashed lines) and the corrections due to the frequency-dependent terms with one and two $\bfD$ operators (dotted and dash-dotted lines, respectively). From Figs.~\ref{fig:1s2p1_0}-\ref{fig:fig:quasi_12} it is seen that there is a cancellation between the different terms, which is especially pronounced for $(1s 2p_{1/2})_{0}$ state and for the off-diagonal matrix element. From Figs.~\ref{fig:1s2p1_0}-\ref{fig:fig:quasi_12} it follows also that in the region $Z\lesssim 40$ the higher-order corrections $\Delta A^{(\rm 2el)}(\alpha Z)$ divided by $(\alpha Z)^4$ are almost constant, excepting the case of the total contribution for the off-diagonal matrix element for which the corresponding term crosses zero. It means that the dominant contribution to $\Delta A^{(\rm 2el)}(\alpha Z)$ comes from the $(\alpha Z)^6 m^2/M$ term,
\be\label{eq:deltaA_expan}
\Delta A^{(\rm 2el)}(\alpha Z) = a_4 (\alpha Z)^4 + \ldots \,.
\ee
The coefficient $a_4$ is known analytically for the term with two $\bfD$ operators in the case of $(1s^2)2p_{1/2}$  and $(1s^2)2p_{3/2}$  states in Li-like ions~\cite{Artemyev:1995:1884,Artemyev:1995:5201}. We note that in Ref.~\cite{Artemyev:1995:5201}, due to a misprint, the coefficient $a_4$ is presented with the opposite sign. For all other higher-order corrections shown in Figs.~\ref{fig:1s2p1_0}-\ref{fig:fig:quasi_12} the coefficients have been obtained by fitting our numerical results. The calculated coefficients for $(1s 2p_{1/2})_0$ and $(1s 2p_{3/2})_2$ states and for the off-diagonal matrix element are presented in Table~\ref{tab:A4_expan}. The corresponding coefficients for $(1s 2p_{1/2})_1$ and $(1s 2p_{3/2})_1$ states and for Li-like ions can be obtained with the use of Eqs.~(\ref{eq:dE_p1_2el_relation}) and (\ref{eq:dE_p3_2el_relation}). We note that even for $Z=50$ the term  $a_4(\alpha Z)^4$  provides the dominant correction. For instance, the total two-electron recoil contribution for $(1s 2p_{1/2})_{0}$ level evaluated using the $\alpha Z$ expansion up to the order $(\alpha Z)^6 m^2/M$ equals to $-0.07285$, while the all-order value is $-0.07282$. Nevertheless, the $\alpha Z$-expansion results play only a supporting role in the present consideration to zeroth order in $1/Z$, since the complete $\alpha Z$-dependent calculation is performed.


The recoil contribution of the first order in $1/Z$ has been calculated by perturbation theory within the lowest-order relativistic approximation using the operator~(\ref{br1}) and the standard expression for the interelectronic-interaction operator valid to this order. This operator being the sum of the Coulomb and Breit electron-electron interaction operators reads as
\begin{eqnarray} \label{br}
V_{ij} &=& e^2\alpha_i^{\rho}\alpha_j^{\sigma}D_{\rho \sigma}(0,{\bf r}_{ij})
= V^{\rm C}_{ij}+ V^{\rm B}_{ij}\nonumber \\
&=&
\frac{\alpha}{r_{ij}} 
- \frac{\alpha}{2}
  \left[
    \frac{\balpha_i \cdot \balpha_j}{r_{ij}} 
         + \frac{(\balpha_i \cdot \bfr_{ij})(\balpha_j \cdot \bfr_{ij})}{r_{ij}^3}
  \right]
\,
\end{eqnarray}
where the indices $i$ and $j$ enumerate the atomic electrons. The summations over intermediate electron states, which are restricted to the positive-energy part of the Dirac spectrum, have been performed using the DKB finite basis set method with the basis functions constructed from B-splines. The results of the calculations, expressed in terms of the function $B(\alpha Z)$, 
\begin{eqnarray}\label{B}
\Delta E = \frac{(\alpha Z)^2}{Z}\frac{m^2}{M}B(\alpha Z)\,,
\end{eqnarray}
are presented in Table~\ref{two-electron_1}. For comparison, we present also the nonrelativistic results corresponding to the limit $B(0)$ and the difference between the relativistic and nonrelativistic data. The nonrelativistic values were obtained using the $1/Z$-expansion coefficients from Refs.~\cite{Drake:1988:586,Sanders:1969:84,Aashamar:1970:3324}.

For the ions under consideration ($Z=12-100$), the recoil corrections of the second and higher orders in $1/Z$ can be taken into account within the nonrelativistic approximation by summing the related contributions to the normal and specific mass shifts. The normal mass shift was calculated as $-(m/M) \Delta E_{\rm 2ph}$, where $\Delta E_{\rm 2ph}$ is the two-photon exchange contribution to the binding energy evaluated within the Breit approximation. The specific mass shift was calculated employing the $1/Z$-expansion coefficients from Ref.~\cite{Drake:1988:586}. These corrections, which are very small, have been added to the total low-order results presented below.

In Table~\ref{He} we present the total values of the nuclear recoil contributions to the binding and ionization energies in He-like ions in the range $Z=12-100$. The results for the ionization energies are obtained by subtracting the recoil contributions for the $1s$ state from the total nuclear recoil contributions to the binding energies. In Table~\ref{He} the QED term includes only the higher-order one-electron contribution. All other contributions, including the frequency-dependent correction to the two-electron recoil part of zeroth order in $1/Z$, are referred as non-QED terms here. The electron-to-nucleus mass ratios were taken from the Ame2012 compilation~\cite{Wang:2012:1603}. The uncertainty is obtained as a sum of two contributions. The first one is due to the approximate treatment of the finite nuclear size effect on the one-electron QED recoil contribution. This uncertainty is estimated as described above. The second uncertainty is due to uncalculated QED recoil corrections of the first order in $1/Z$. It was estimated as a sum of the one-electron QED correction divided by $Z$ and the relativistic recoil correction of the $1/Z$ order (the last column in Table~\ref{two-electron_1}) multiplied by a factor $\alpha Z$. For comparison, we give also the values of the recoil effect on the ionization energies obtained in Ref.~\cite{Artemyev:2005:062104}. Compared to Ref.~\cite{Artemyev:2005:062104}, our calculations include a more accurate treatment of the one-electron recoil contribution, the complete $\alpha Z$-dependent calculation of the two-electron recoil contribution of zeroth order in $1/Z$, and the calculation of the relativistic correction to the $1/Z$ recoil term. In contrast to Ref.~\cite{Artemyev:2005:062104}, we indicate explicitly the uncertainties of the total results. It can be seen that in the region $Z\le 40$, where the systematic discrepancy  between theory and experiment was announced~\cite{Chantler:2012:153001}, the obtained results coincide with those from Ref.~\cite{Artemyev:2005:062104} almost to all indicated digits. Some differences between the present results and those from Ref.~\cite{Artemyev:2005:062104}, which occur for heavy ions, are much smaller than the total theoretical uncertainties of the ionization energies given in Ref.~\cite{Artemyev:2005:062104}.

\section{Conclusion}
In this paper we have evaluated the nuclear recoil effect on the binding energies in highly charged He-like ions. The calculations included the $m/M$ one-electron recoil contribution in the framework of the rigorous QED formalism and the two-electron recoil contribution within the framework of the Breit approximation, with taking into account the frequency-dependent correction to zeroth order in $1/Z$. The estimated uncertainties are due to  the approximate treatment of the finite nuclear size correction to the QED recoil effect and uncalculated QED recoil corrections of the first order in $1/Z$. The obtained results agree with the previous calculations~\cite{Artemyev:2005:062104} but have a better accuracy. Thus, the recoil effect can not be responsible for the discrepancy  between theory and experiment claimed in Ref.~\cite{Chantler:2012:153001}.


\section{Acknowledgments}

This work was supported by the Russian Science Foundation (Grant No. 17-12-01097).


\input{tables_new_style.tex}



\end{document}

%% file: tables_new_style.tex
\begin{longtable}{
                 S[table-format=3.0,group-separator=]
                 S[table-format=3.4(2),group-separator=]@{\quad\,\,}
                 l@{\quad\,}
                 S[table-format=-1.5,group-separator=]@{\quad\,\,}
                 S[table-format=-1.5,group-separator=]@{\quad\,\,}
                 S[table-format=-1.5,group-separator=]@{\quad\,\,}
                 S[table-format=-1.5(2),group-separator=]
                 }
  \caption{\label{H-like} 
        The one-electron recoil contribution to the binding energy, 
        expressed in terms of the function $A(\alpha Z)$ defined by 
        equation (\ref{A}). The terms $A_{\rm L}(\alpha Z)$ and $A_{\rm H}(\alpha Z)$ correspond to the 
        low- and higher-order contributions evaluated for extended nuclei by Eqs. (\ref{low}) and (\ref{high}), respectively. 
        The term $\delta A_{\rm L}^{\rm (NS,add)}$ gives the additional nuclear size correction to the 
        low-order contribution, which is beyond that accounted for by formula (\ref{low}) 
        with the potential of an extended nucleus. 
        $A_{\rm tot}(\alpha Z)$ is the sum of all the contributions.}\\
\hline
 \multicolumn{1}{r}{$Z$} & \multicolumn{1}{c}{$\langle r^2 \rangle^{1/2}$~[fm]} & \multicolumn{1}{l}{State} & 
 \multicolumn{1}{c}{$A_{\rm L}(\alpha Z)$\,} & \multicolumn{1}{c}{$A_{\rm H}(\alpha Z)$\,} & 
 \multicolumn{1}{c}{$\delta A_{\rm L}^{\rm (NS,add)}$} & \multicolumn{1}{c}{$A_{\rm tot}(\alpha Z)$\,\,} \\
\hline
\endfirsthead
\caption{\it (Continued.)}\\
\hline
 \multicolumn{1}{r}{$Z$} & \multicolumn{1}{c}{$\langle r^2 \rangle^{1/2}$~[fm]} & \multicolumn{1}{l}{State} & 
 \multicolumn{1}{c}{$A_{\rm L}(\alpha Z)$\,} & \multicolumn{1}{c}{$A_{\rm H}(\alpha Z)$\,} & 
 \multicolumn{1}{c}{$\delta A_{\rm L}^{\rm (NS,add)}$} & \multicolumn{1}{c}{$A_{\rm tot}(\alpha Z)$\,\,} \\
\hline
\endhead
\hline
\endfoot
\hline
\endlastfoot
                       
   12  &   3.0570(16)  &  $      1s$ &    0.49999 &    0.00079 &    0.00000 &       0.50078    \\ 
       &               &  $      2s$ &    0.12524 &    0.00012 &    0.00000 &       0.12536    \\ 
       &               &  $2p_{1/2}$ &    0.12524 &   -0.00001 &    0.00000 &       0.12523    \\ 
       &               &  $2p_{3/2}$ &    0.12500 &   -0.00001 &    0.00000 &       0.12499    \\ 

\hline
   20  &   3.4776(19)  &  $      1s$ &    0.49995 &    0.00329 &    0.00000 &       0.50325    \\ 
       &               &  $      2s$ &    0.12567 &    0.00051 &    0.00000 &       0.12618    \\ 
       &               &  $2p_{1/2}$ &    0.12567 &   -0.00002 &    0.00000 &       0.12566    \\ 
       &               &  $2p_{3/2}$ &    0.12500 &   -0.00002 &    0.00000 &       0.12498    \\ 

\hline
   30  &   3.9283(15)  &  $      1s$ &    0.49980 &    0.01031 &    0.00001 &       0.51011    \\ 
       &               &  $      2s$ &    0.12651 &    0.00165 &    0.00000 &       0.12816    \\ 
       &               &  $2p_{1/2}$ &    0.12653 &   -0.00002 &    0.00000 &       0.12652    \\ 
       &               &  $2p_{3/2}$ &    0.12500 &   -0.00005 &    0.00000 &       0.12495    \\ 

\hline
   40  &   4.2694(10)  &  $      1s$ &    0.49937 &    0.02357 &    0.00003 &       0.52297    \\ 
       &               &  $      2s$ &    0.12770 &    0.00390 &    0.00000 &       0.13160    \\ 
       &               &  $2p_{1/2}$ &    0.12778 &    0.00007 &    0.00000 &       0.12785    \\ 
       &               &  $2p_{3/2}$ &    0.12500 &   -0.00005 &    0.00000 &       0.12495    \\ 

\hline
   50  &   4.6519(21)  &  $      1s$ &    0.49831 &    0.04597 &    0.00011 &    0.54439(1)    \\ 
       &               &  $      2s$ &    0.12922 &    0.00790 &    0.00002 &       0.13714    \\ 
       &               &  $2p_{1/2}$ &    0.12946 &    0.00039 &    0.00000 &       0.12985    \\ 
       &               &  $2p_{3/2}$ &    0.12500 &    0.00001 &    0.00000 &       0.12501    \\ 

\hline
   60  &   4.9123(25)  &  $      1s$ &    0.49590 &    0.08207 &    0.00037 &    0.57834(6)    \\ 
       &               &  $      2s$ &    0.13103 &    0.01470 &    0.00006 &    0.14579(1)    \\ 
       &               &  $2p_{1/2}$ &    0.13162 &    0.00123 &    0.00000 &       0.13285    \\ 
       &               &  $2p_{3/2}$ &    0.12500 &    0.00020 &    0.00000 &       0.12520    \\ 

\hline
   70  &   5.3108(60)  &  $      1s$ &    0.49029 &    0.13979 &    0.00111 &   0.63119(32)    \\ 
       &               &  $      2s$ &    0.13289 &    0.02620 &    0.00018 &    0.15927(4)    \\ 
       &               &  $2p_{1/2}$ &    0.13434 &    0.00315 &    0.00001 &       0.13751    \\ 
       &               &  $2p_{3/2}$ &    0.12500 &    0.00060 &    0.00000 &       0.12560    \\ 

\hline
   80  &   5.4648(33)  &  $      1s$ &     0.4782 &     0.2334 &     0.0030 &    0.7147(15)    \\ 
       &               &  $      2s$ &    0.13424 &    0.04606 &    0.00053 &   0.18083(18)    \\ 
       &               &  $2p_{1/2}$ &    0.13767 &    0.00732 &    0.00005 &       0.14504    \\ 
       &               &  $2p_{3/2}$ &    0.12500 &    0.00130 &    0.00000 &       0.12630    \\ 

\hline
   90  &    5.785(12)  &  $      1s$ &     0.4505 &     0.3905 &     0.0081 &    0.8491(70)    \\ 
       &               &  $      2s$ &    0.13323 &    0.08179 &    0.00154 &   0.21657(95)    \\ 
       &               &  $2p_{1/2}$ &    0.14147 &    0.01642 &    0.00018 &    0.15807(2)    \\ 
       &               &  $2p_{3/2}$ &    0.12500 &    0.00240 &    0.00000 &       0.12740    \\ 

\hline
   92  &   5.8571(33)  &  $      1s$ &     0.4416 &     0.4336 &     0.0099 &    0.8851(97)    \\ 
       &               &  $      2s$ &     0.1324 &     0.0920 &     0.0019 &    0.2263(13)    \\ 
       &               &  $2p_{1/2}$ &    0.14223 &    0.01929 &    0.00024 &    0.16177(3)    \\ 
       &               &  $2p_{3/2}$ &    0.12500 &    0.00269 &    0.00000 &       0.12768    \\ 

\hline
  100  &    5.857(59)  &  $      1s$ &      0.388 &      0.669 &      0.021 &     1.078(36)    \\ 
       &               &  $      2s$ &     0.1253 &     0.1502 &     0.0044 &    0.2800(53)    \\ 
       &               &  $2p_{1/2}$ &    0.14489 &    0.03731 &    0.00071 &   0.18291(19)    \\ 
       &               &  $2p_{3/2}$ &    0.12500 &    0.00404 &    0.00000 &       0.12904    \\ 

\hline

\end{longtable}


\begin{longtable}{
                 S[table-format=3.2,group-separator=]
                 l
                 S[table-format=-2.6,group-separator=]
                 S[table-format=-2.6,group-separator=]
                 S[table-format=-2.6,group-separator=]
                 S[table-format=-2.6,group-separator=]
                 }
  \caption{\label{two-electron_0} 
        The two-electron recoil contribution of zeroth order in $1/Z$, 
        expressed in terms of the function $A(\alpha Z)$ defined by equation (\ref{A}). 
        The frequency-dependent correction includes also the term with two $\bfD$ operators 
        in Eqs.~(\ref{two}) and (\ref{eq:2el_rec_quasi}). }\\
\hline
 \multicolumn{1}{c}{$Z$}  &  \multicolumn{1}{c}{State}  &  \multicolumn{1}{c}{Breit at zero}    &  
 \multicolumn{1}{c}{Freq.-depend.} & \multicolumn{1}{c}{Total,}    & \multicolumn{1}{c}{$\alpha Z$-expansion} \\[-1mm]
                           &                             &  \multicolumn{1}{c}{energy transfer}  &  
 \multicolumn{1}{c}{correction}    & \multicolumn{1}{c}{this work} & \multicolumn{1}{c}{formula, Ref. \cite{Shabaev:1994:1307}}      \\        
\hline
\endfirsthead
\caption{\it (Continued.)}\\
\hline
 \multicolumn{1}{c}{$Z$}  &  \multicolumn{1}{c}{State}  &  \multicolumn{1}{c}{Breit at zero}    &  
 \multicolumn{1}{c}{Freq.-depend.} & \multicolumn{1}{c}{Total,}    & \multicolumn{1}{c}{$\alpha Z$-expansion} \\[-1mm]
                           &                             &  \multicolumn{1}{c}{energy transfer}  &  
 \multicolumn{1}{c}{correction}    & \multicolumn{1}{c}{this work} & \multicolumn{1}{c}{formula, Ref. \cite{Shabaev:1994:1307}}      \\    
\hline
\endhead
\hline
\endfoot
\hline
\endlastfoot
                       
   12  &  $(1s 2p_{1/2})_0$  &    -0.07774   &     0.00000   &    -0.07775   &    -0.07775     \\ 
       &  $(1s 2p_{1/2})_1$  &    -0.02591   &     0.00000   &    -0.02592   &    -0.02592     \\ 
       &  $(1s 2p_{3/2})_1$  &     0.02594   &     0.00000   &     0.02594   &     0.02594     \\ 
       &  $(1s 2p_{3/2})_2$  &    -0.07782   &     0.00000   &    -0.07782   &    -0.07782     \\ 
       &   {\rm off-diag.}   &     0.07333   &     0.00000   &     0.07334   &     0.07334     \\ 

\hline
   20  &  $(1s 2p_{1/2})_0$  &    -0.07720   &    -0.00003   &    -0.07722   &    -0.07723     \\ 
       &  $(1s 2p_{1/2})_1$  &    -0.02573   &    -0.00001   &    -0.02574   &    -0.02574     \\ 
       &  $(1s 2p_{3/2})_1$  &     0.02581   &     0.00000   &     0.02582   &     0.02582     \\ 
       &  $(1s 2p_{3/2})_2$  &    -0.07744   &    -0.00001   &    -0.07745   &    -0.07745     \\ 
       &   {\rm off-diag.}   &     0.07290   &     0.00001   &     0.07291   &     0.07291     \\ 

\hline
   30  &  $(1s 2p_{1/2})_0$  &    -0.07605   &    -0.00015   &    -0.07620   &    -0.07622     \\ 
       &  $(1s 2p_{1/2})_1$  &    -0.02535   &    -0.00005   &    -0.02540   &    -0.02541     \\ 
       &  $(1s 2p_{3/2})_1$  &     0.02556   &     0.00002   &     0.02558   &     0.02557     \\ 
       &  $(1s 2p_{3/2})_2$  &    -0.07667   &    -0.00006   &    -0.07673   &    -0.07671     \\ 
       &   {\rm off-diag.}   &     0.07202   &     0.00007   &     0.07209   &     0.07209     \\ 

\hline
   40  &  $(1s 2p_{1/2})_0$  &    -0.07423   &    -0.00051   &    -0.07474   &    -0.07480     \\ 
       &  $(1s 2p_{1/2})_1$  &    -0.02474   &    -0.00017   &    -0.02491   &    -0.02493     \\ 
       &  $(1s 2p_{3/2})_1$  &     0.02518   &     0.00007   &     0.02525   &     0.02522     \\ 
       &  $(1s 2p_{3/2})_2$  &    -0.07554   &    -0.00020   &    -0.07574   &    -0.07567     \\ 
       &   {\rm off-diag.}   &     0.07070   &     0.00023   &     0.07094   &     0.07093     \\ 

\hline
   50  &  $(1s 2p_{1/2})_0$  &    -0.07151   &    -0.00131   &    -0.07282   &    -0.07298     \\ 
       &  $(1s 2p_{1/2})_1$  &    -0.02384   &    -0.00044   &    -0.02427   &    -0.02433     \\ 
       &  $(1s 2p_{3/2})_1$  &     0.02467   &     0.00016   &     0.02483   &     0.02478     \\ 
       &  $(1s 2p_{3/2})_2$  &    -0.07401   &    -0.00049   &    -0.07450   &    -0.07434     \\ 
       &   {\rm off-diag.}   &     0.06886   &     0.00058   &     0.06944   &     0.06945     \\ 

\hline
   60  &  $(1s 2p_{1/2})_0$  &    -0.06744   &    -0.00295   &    -0.07039   &    -0.07076     \\ 
       &  $(1s 2p_{1/2})_1$  &    -0.02248   &    -0.00098   &    -0.02346   &    -0.02359     \\ 
       &  $(1s 2p_{3/2})_1$  &     0.02400   &     0.00034   &     0.02434   &     0.02424     \\ 
       &  $(1s 2p_{3/2})_2$  &    -0.07201   &    -0.00102   &    -0.07302   &    -0.07271     \\ 
       &   {\rm off-diag.}   &     0.06635   &     0.00124   &     0.06759   &     0.06763     \\ 

\hline
   70  &  $(1s 2p_{1/2})_0$  &    -0.06133   &    -0.00604   &    -0.06737   &    -0.06813     \\ 
       &  $(1s 2p_{1/2})_1$  &    -0.02044   &    -0.00201   &    -0.02246   &    -0.02271     \\ 
       &  $(1s 2p_{3/2})_1$  &     0.02314   &     0.00063   &     0.02377   &     0.02360     \\ 
       &  $(1s 2p_{3/2})_2$  &    -0.06943   &    -0.00188   &    -0.07131   &    -0.07079     \\ 
       &   {\rm off-diag.}   &     0.06295   &     0.00238   &     0.06534   &     0.06549     \\ 

\hline
   80  &  $(1s 2p_{1/2})_0$  &    -0.05189   &    -0.01175   &    -0.06363   &    -0.06510     \\ 
       &  $(1s 2p_{1/2})_1$  &    -0.01730   &    -0.00392   &    -0.02121   &    -0.02170     \\ 
       &  $(1s 2p_{3/2})_1$  &     0.02205   &     0.00107   &     0.02312   &     0.02286     \\ 
       &  $(1s 2p_{3/2})_2$  &    -0.06614   &    -0.00321   &    -0.06936   &    -0.06857     \\ 
       &   {\rm off-diag.}   &     0.05835   &     0.00426   &     0.06261   &     0.06302     \\ 

\hline
   90  &  $(1s 2p_{1/2})_0$  &    -0.03669   &    -0.02229   &    -0.05899   &    -0.06167     \\ 
       &  $(1s 2p_{1/2})_1$  &    -0.01223   &    -0.00743   &    -0.01966   &    -0.02056     \\ 
       &  $(1s 2p_{3/2})_1$  &     0.02065   &     0.00172   &     0.02237   &     0.02202     \\ 
       &  $(1s 2p_{3/2})_2$  &    -0.06196   &    -0.00516   &    -0.06712   &    -0.06606     \\ 
       &   {\rm off-diag.}   &     0.05199   &     0.00728   &     0.05927   &     0.06021     \\ 

\hline
   92  &  $(1s 2p_{1/2})_0$  &    -0.03260   &    -0.02533   &    -0.05793   &    -0.06093     \\ 
       &  $(1s 2p_{1/2})_1$  &    -0.01087   &    -0.00844   &    -0.01931   &    -0.02031     \\ 
       &  $(1s 2p_{3/2})_1$  &     0.02033   &     0.00188   &     0.02221   &     0.02184     \\ 
       &  $(1s 2p_{3/2})_2$  &    -0.06100   &    -0.00563   &    -0.06663   &    -0.06552     \\ 
       &   {\rm off-diag.}   &     0.05044   &     0.00807   &     0.05851   &     0.05961     \\ 

\hline
  100  &  $(1s 2p_{1/2})_0$  &    -0.01065   &    -0.04248   &    -0.05312   &    -0.05782     \\ 
       &  $(1s 2p_{1/2})_1$  &    -0.00355   &    -0.01416   &    -0.01771   &    -0.01927     \\ 
       &  $(1s 2p_{3/2})_1$  &     0.01887   &     0.00262   &     0.02150   &     0.02108     \\ 
       &  $(1s 2p_{3/2})_2$  &    -0.05662   &    -0.00787   &    -0.06449   &    -0.06325     \\ 
       &   {\rm off-diag.}   &     0.04295   &     0.01213   &     0.05508   &     0.05708     \\ 

\hline

\end{longtable}


\begin{longtable}{
                 l@{\quad}
                 S[table-format=-2.5(2),group-separator=]@{\quad\quad}
                 S[table-format=-2.5(1),group-separator=]@{\quad\quad}
                 S[table-format=-2.5(1),group-separator=]@{\quad\quad}
                 S[table-format=-2.5(2),group-separator=]
                 }
\caption{\label{tab:A4_expan} 
\label{tab:expan} The coefficients $a_4$ defined by Eq.~(\ref{eq:deltaA_expan}) for
the two-electron recoil contribution to the binding energies in He-like ions to zeroth order in $1/Z$.}\\
\hline
 \multicolumn{1}{l}{\quad State} & \multicolumn{1}{c}{Breit at zero}   & \multicolumn{1}{c}{Freq.-depend.}        & 
         \multicolumn{1}{c}{Freq.-depend.}        & \multicolumn{1}{c}{Total} \\[-1mm]
       & \multicolumn{1}{c}{energy transfer} & \multicolumn{1}{c}{term with one $\bfD$} & 
         \multicolumn{1}{c}{term with two $\bfD$} &       \\
\hline
\endfirsthead
\caption{\it (Continued.)}\\
\hline
 \multicolumn{1}{l}{\quad State} & \multicolumn{1}{c}{Breit at zero}   & \multicolumn{1}{c}{Freq.-depend.}        & 
         \multicolumn{1}{c}{Freq.-depend.}        & \multicolumn{1}{c}{Total} \\[-1mm]
       & \multicolumn{1}{c}{energy transfer} & \multicolumn{1}{c}{term with one $\bfD$} & 
         \multicolumn{1}{c}{term with two $\bfD$} &       \\
\hline
\endhead
\hline
\endfoot
\hline
\endlastfoot

\\[-6mm]
   
 $(1s2p_{1/2})_0$    &  
  0.07086(10)           &  
 -0.00366          &                     
 -0.05975           &     
  0.00746(10)            \\[3mm]            

 $(1s2p_{3/2})_2$    &  
  0.01755(5)          &  
 -0.02012(1)          &  
 -0.00762           &     
 -0.01019(5)          \\[3mm]

 off-diagonal        &  
 -0.03019(7)          &  
  0.01121(1)           &              
  0.02012(1)           &              
  0.00113(6)            \\[3mm]

\hline

\end{longtable}


\begin{longtable}{
                 S[table-format=3.2,group-separator=]
                 l
                 S[table-format=-5.7,group-separator=]
                 S[table-format=-3.7,group-separator=]
                 S[table-format=-3.7,group-separator=]
                 }
 \caption{\label{two-electron_1} 
       The recoil contribution of first order  in $1/Z$, 
       expressed in terms of the function $B(\alpha Z)$ defined by equation (\ref{B}).}\\
\hline
  \multicolumn{1}{c}{$Z$\,}  &   \multicolumn{1}{l}{State}  &  \multicolumn{1}{c}{Relativistic}  &  
  \multicolumn{1}{c}{Nonrelativistic}  &  \multicolumn{1}{c}{Difference} \\     
\hline
\endfirsthead
\caption{\it (Continued.)}\\
\hline
  \multicolumn{1}{c}{$Z$\,}  &   \multicolumn{1}{l}{State}  &  \multicolumn{1}{c}{Relativistic}  &  
  \multicolumn{1}{c}{Nonrelativistic}  &  \multicolumn{1}{c}{Difference} \\     
\hline
\endhead
\hline
\endfoot
\hline
\endlastfoot
                       
   12  &           $(1s)^2$  &     -0.4967   &     -0.4917   &     -0.0050     \\ 
       &        $(1s 2s)_0$  &     -0.2211   &     -0.2194   &     -0.0017     \\ 
       &        $(1s 2s)_1$  &     -0.1776   &     -0.1770   &     -0.0006     \\ 
       &  $(1s 2p_{1/2})_0$  &     -0.0843   &     -0.0826   &     -0.0017     \\ 
       &  $(1s 2p_{1/2})_1$  &     -0.1966   &     -0.1958   &     -0.0008     \\ 
       &  $(1s 2p_{3/2})_1$  &     -0.3092   &     -0.3090   &     -0.0002     \\ 
       &  $(1s 2p_{3/2})_2$  &     -0.0828   &     -0.0826   &     -0.0002     \\ 
       &   {\rm off-diag.}   &     -0.1597   &     -0.1601   &      0.0004     \\ 

\hline
   20  &           $(1s)^2$  &     -0.5055   &     -0.4917   &     -0.0138     \\ 
       &        $(1s 2s)_0$  &     -0.2237   &     -0.2194   &     -0.0043     \\ 
       &        $(1s 2s)_1$  &     -0.1787   &     -0.1770   &     -0.0017     \\ 
       &  $(1s 2p_{1/2})_0$  &     -0.0874   &     -0.0826   &     -0.0048     \\ 
       &  $(1s 2p_{1/2})_1$  &     -0.1980   &     -0.1958   &     -0.0022     \\ 
       &  $(1s 2p_{3/2})_1$  &     -0.3095   &     -0.3090   &     -0.0005     \\ 
       &  $(1s 2p_{3/2})_2$  &     -0.0831   &     -0.0826   &     -0.0006     \\ 
       &   {\rm off-diag.}   &     -0.1589   &     -0.1601   &      0.0012     \\ 

\hline
   30  &           $(1s)^2$  &     -0.5230   &     -0.4917   &     -0.0313     \\ 
       &        $(1s 2s)_0$  &     -0.2289   &     -0.2194   &     -0.0096     \\ 
       &        $(1s 2s)_1$  &     -0.1808   &     -0.1770   &     -0.0038     \\ 
       &  $(1s 2p_{1/2})_0$  &     -0.0940   &     -0.0826   &     -0.0115     \\ 
       &  $(1s 2p_{1/2})_1$  &     -0.2010   &     -0.1958   &     -0.0052     \\ 
       &  $(1s 2p_{3/2})_1$  &     -0.3102   &     -0.3090   &     -0.0012     \\ 
       &  $(1s 2p_{3/2})_2$  &     -0.0839   &     -0.0826   &     -0.0013     \\ 
       &   {\rm off-diag.}   &     -0.1574   &     -0.1601   &      0.0028     \\ 

\hline
   40  &           $(1s)^2$  &     -0.5478   &     -0.4917   &     -0.0561     \\ 
       &        $(1s 2s)_0$  &     -0.2369   &     -0.2194   &     -0.0175     \\ 
       &        $(1s 2s)_1$  &     -0.1839   &     -0.1770   &     -0.0069     \\ 
       &  $(1s 2p_{1/2})_0$  &     -0.1046   &     -0.0826   &     -0.0220     \\ 
       &  $(1s 2p_{1/2})_1$  &     -0.2056   &     -0.1958   &     -0.0098     \\ 
       &  $(1s 2p_{3/2})_1$  &     -0.3112   &     -0.3090   &     -0.0022     \\ 
       &  $(1s 2p_{3/2})_2$  &     -0.0850   &     -0.0826   &     -0.0025     \\ 
       &   {\rm off-diag.}   &     -0.1550   &     -0.1601   &      0.0052     \\ 

\hline
   50  &           $(1s)^2$  &     -0.5798   &     -0.4917   &     -0.0881     \\ 
       &        $(1s 2s)_0$  &     -0.2479   &     -0.2194   &     -0.0286     \\ 
       &        $(1s 2s)_1$  &     -0.1880   &     -0.1770   &     -0.0110     \\ 
       &  $(1s 2p_{1/2})_0$  &     -0.1207   &     -0.0826   &     -0.0382     \\ 
       &  $(1s 2p_{1/2})_1$  &     -0.2124   &     -0.1958   &     -0.0166     \\ 
       &  $(1s 2p_{3/2})_1$  &     -0.3127   &     -0.3090   &     -0.0037     \\ 
       &  $(1s 2p_{3/2})_2$  &     -0.0867   &     -0.0826   &     -0.0041     \\ 
       &   {\rm off-diag.}   &     -0.1515   &     -0.1601   &      0.0086     \\ 

\hline
   60  &           $(1s)^2$  &     -0.6187   &     -0.4917   &     -0.1270     \\ 
       &        $(1s 2s)_0$  &     -0.2628   &     -0.2194   &     -0.0434     \\ 
       &        $(1s 2s)_1$  &     -0.1930   &     -0.1770   &     -0.0160     \\ 
       &  $(1s 2p_{1/2})_0$  &     -0.1451   &     -0.0826   &     -0.0625     \\ 
       &  $(1s 2p_{1/2})_1$  &     -0.2222   &     -0.1958   &     -0.0264     \\ 
       &  $(1s 2p_{3/2})_1$  &     -0.3146   &     -0.3090   &     -0.0056     \\ 
       &  $(1s 2p_{3/2})_2$  &     -0.0890   &     -0.0826   &     -0.0064     \\ 
       &   {\rm off-diag.}   &     -0.1465   &     -0.1601   &      0.0136     \\ 

\hline
   70  &           $(1s)^2$  &     -0.6618   &     -0.4917   &     -0.1701     \\ 
       &        $(1s 2s)_0$  &     -0.2819   &     -0.2194   &     -0.0625     \\ 
       &        $(1s 2s)_1$  &     -0.1986   &     -0.1770   &     -0.0216     \\ 
       &  $(1s 2p_{1/2})_0$  &     -0.1822   &     -0.0826   &     -0.0996     \\ 
       &  $(1s 2p_{1/2})_1$  &     -0.2365   &     -0.1958   &     -0.0407     \\ 
       &  $(1s 2p_{3/2})_1$  &     -0.3170   &     -0.3090   &     -0.0080     \\ 
       &  $(1s 2p_{3/2})_2$  &     -0.0920   &     -0.0826   &     -0.0095     \\ 
       &   {\rm off-diag.}   &     -0.1395   &     -0.1601   &      0.0206     \\ 

\hline
   80  &           $(1s)^2$  &     -0.7030   &     -0.4917   &     -0.2113     \\ 
       &        $(1s 2s)_0$  &     -0.3049   &     -0.2194   &     -0.0855     \\ 
       &        $(1s 2s)_1$  &     -0.2040   &     -0.1770   &     -0.0270     \\ 
       &  $(1s 2p_{1/2})_0$  &     -0.2400   &     -0.0826   &     -0.1575     \\ 
       &  $(1s 2p_{1/2})_1$  &     -0.2573   &     -0.1958   &     -0.0615     \\ 
       &  $(1s 2p_{3/2})_1$  &     -0.3200   &     -0.3090   &     -0.0110     \\ 
       &  $(1s 2p_{3/2})_2$  &     -0.0961   &     -0.0826   &     -0.0135     \\ 
       &   {\rm off-diag.}   &     -0.1295   &     -0.1601   &      0.0307     \\ 

\hline
   90  &           $(1s)^2$  &     -0.7196   &     -0.4917   &     -0.2279     \\ 
       &        $(1s 2s)_0$  &     -0.3264   &     -0.2194   &     -0.1070     \\ 
       &        $(1s 2s)_1$  &     -0.2056   &     -0.1770   &     -0.0286     \\ 
       &  $(1s 2p_{1/2})_0$  &     -0.3324   &     -0.0826   &     -0.2498     \\ 
       &  $(1s 2p_{1/2})_1$  &     -0.2873   &     -0.1958   &     -0.0915     \\ 
       &  $(1s 2p_{3/2})_1$  &     -0.3235   &     -0.3090   &     -0.0145     \\ 
       &  $(1s 2p_{3/2})_2$  &     -0.1010   &     -0.0826   &     -0.0185     \\ 
       &   {\rm off-diag.}   &     -0.1146   &     -0.1601   &      0.0455     \\ 

\hline
   92  &           $(1s)^2$  &     -0.7158   &     -0.4917   &     -0.2242     \\ 
       &        $(1s 2s)_0$  &     -0.3292   &     -0.2194   &     -0.1099     \\ 
       &        $(1s 2s)_1$  &     -0.2048   &     -0.1770   &     -0.0278     \\ 
       &  $(1s 2p_{1/2})_0$  &     -0.3569   &     -0.0826   &     -0.2743     \\ 
       &  $(1s 2p_{1/2})_1$  &     -0.2946   &     -0.1958   &     -0.0988     \\ 
       &  $(1s 2p_{3/2})_1$  &     -0.3242   &     -0.3090   &     -0.0152     \\ 
       &  $(1s 2p_{3/2})_2$  &     -0.1021   &     -0.0826   &     -0.0196     \\ 
       &   {\rm off-diag.}   &     -0.1109   &     -0.1601   &      0.0492     \\ 

\hline
  100  &           $(1s)^2$  &     -0.6556   &     -0.4917   &     -0.1639     \\ 
       &        $(1s 2s)_0$  &     -0.3277   &     -0.2194   &     -0.1083     \\ 
       &        $(1s 2s)_1$  &     -0.1943   &     -0.1770   &     -0.0173     \\ 
       &  $(1s 2p_{1/2})_0$  &     -0.4850   &     -0.0826   &     -0.4024     \\ 
       &  $(1s 2p_{1/2})_1$  &     -0.3286   &     -0.1958   &     -0.1329     \\ 
       &  $(1s 2p_{3/2})_1$  &     -0.3275   &     -0.3090   &     -0.0185     \\ 
       &  $(1s 2p_{3/2})_2$  &     -0.1069   &     -0.0826   &     -0.0244     \\ 
       &   {\rm off-diag.}   &     -0.0919   &     -0.1601   &      0.0682     \\ 

\hline

\end{longtable}


\begin{longtable}{
                 S[table-format=3.2,group-separator=]
                 l
                 S[table-format=-4.5,group-separator=]
                 S[table-format=-4.5,group-separator=]
                 S[table-format=-4.6(2),group-separator=]
                 S[table-format=-4.6(2),group-separator=]
                 S[table-format=-3.6,group-separator=]
                 }
 \caption{\label{He} 
      The recoil contributions to the binding and ionization energies    
      in highly charged He-like ions, in eV.}\\
\hline
 \multicolumn{1}{c}{$Z$\,\,\,\,}  &   \multicolumn{1}{l}{State} & \multicolumn{1}{c}{\quad\, Non-QED}  &  \multicolumn{1}{c}{\quad\, QED}   &  
 \multicolumn{1}{c}{Total energy,} &  \multicolumn{1}{c}{Ionization energy,} & \multicolumn{1}{c}{Ionization energy,} \\
                           &                             &                              &                            &  
 \multicolumn{1}{c}{this work}     &  \multicolumn{1}{c}{this work}          & \multicolumn{1}{c}{Ref.~\cite{Artemyev:2005:062104}}  \\
\hline
\endfirsthead
\caption{\it (Continued.)}\\
\hline
 \multicolumn{1}{c}{$Z$\,\,\,\,}  &   \multicolumn{1}{l}{State} & \multicolumn{1}{c}{\quad\, Non-QED}  &  \multicolumn{1}{c}{\quad\, QED}   &  
 \multicolumn{1}{c}{Total energy,} &  \multicolumn{1}{c}{Ionization energy,} & \multicolumn{1}{c}{Ionization energy,} \\
                           &                             &                              &                            &  
 \multicolumn{1}{c}{this work}     &  \multicolumn{1}{c}{this work}          & \multicolumn{1}{c}{Ref.~\cite{Artemyev:2005:062104}}  \\ 
\hline
\endhead
\hline
\endfoot
\hline
\endlastfoot
                       
   12  &           $(1s)^2$  &      0.0860   &      0.0001   &         0.0861   &         0.0412   &      0.0412     \\ 
       &        $(1s 2s)_0$  &      0.0545   &      0.0001   &         0.0545   &         0.0096   &      0.0096     \\ 
       &        $(1s 2s)_1$  &      0.0547   &      0.0001   &         0.0548   &         0.0099   &      0.0099     \\ 
       &  $(1s 2p_{1/2})_0$  &      0.0485   &      0.0001   &         0.0486   &         0.0037   &      0.0037     \\ 
       &  $(1s 2p_{1/2})_1$  &      0.0523   &      0.0001   &         0.0524   &         0.0075   &      0.0075     \\ 
       &  $(1s 2p_{3/2})_1$  &      0.0561   &      0.0001   &         0.0562   &         0.0113   &      0.0113     \\ 
       &  $(1s 2p_{3/2})_2$  &      0.0485   &      0.0001   &         0.0485   &         0.0036   &      0.0036     \\ 
       &   {\rm off-diag.}   &      0.0054   &      0.0000   &         0.0054   &         0.0054   &      0.0054     \\ 

\hline
   20  &           $(1s)^2$  &      0.1457   &      0.0010   &      0.1467(1)   &         0.0715   &      0.0715     \\ 
       &        $(1s 2s)_0$  &      0.0919   &      0.0006   &         0.0924   &         0.0172   &      0.0172     \\ 
       &        $(1s 2s)_1$  &      0.0922   &      0.0006   &         0.0927   &         0.0175   &      0.0175     \\ 
       &  $(1s 2p_{1/2})_0$  &      0.0813   &      0.0005   &         0.0818   &         0.0066   &      0.0066     \\ 
       &  $(1s 2p_{1/2})_1$  &      0.0882   &      0.0005   &         0.0887   &         0.0135   &      0.0135     \\ 
       &  $(1s 2p_{3/2})_1$  &      0.0950   &      0.0005   &         0.0955   &         0.0203   &      0.0203     \\ 
       &  $(1s 2p_{3/2})_2$  &      0.0812   &      0.0005   &         0.0817   &         0.0065   &      0.0065     \\ 
       &   {\rm off-diag.}   &      0.0097   &      0.0000   &         0.0097   &         0.0097   &      0.0097     \\ 

\hline
   30  &           $(1s)^2$  &      0.2065   &      0.0043   &      0.2108(2)   &      0.1036(1)   &      0.1036     \\ 
       &        $(1s 2s)_0$  &      0.1301   &      0.0025   &      0.1326(1)   &         0.0254   &      0.0254     \\ 
       &        $(1s 2s)_1$  &      0.1304   &      0.0025   &      0.1329(1)   &         0.0257   &      0.0257     \\ 
       &  $(1s 2p_{1/2})_0$  &      0.1150   &      0.0022   &      0.1172(1)   &         0.0099   &      0.0099     \\ 
       &  $(1s 2p_{1/2})_1$  &      0.1249   &      0.0022   &      0.1271(1)   &         0.0199   &      0.0199     \\ 
       &  $(1s 2p_{3/2})_1$  &      0.1346   &      0.0022   &      0.1367(1)   &         0.0295   &      0.0295     \\ 
       &  $(1s 2p_{3/2})_2$  &      0.1146   &      0.0022   &      0.1168(1)   &         0.0096   &      0.0096     \\ 
       &   {\rm off-diag.}   &      0.0141   &      0.0000   &         0.0141   &         0.0141   &      0.0140     \\ 

\hline
   40  &           $(1s)^2$  &      0.2618   &      0.0125   &      0.2743(3)   &      0.1353(2)   &      0.1354     \\ 
       &        $(1s 2s)_0$  &      0.1651   &      0.0073   &      0.1724(2)   &         0.0334   &      0.0334     \\ 
       &        $(1s 2s)_1$  &      0.1654   &      0.0073   &      0.1727(2)   &         0.0338   &      0.0338     \\ 
       &  $(1s 2p_{1/2})_0$  &      0.1461   &      0.0063   &      0.1524(2)   &         0.0134   &      0.0134     \\ 
       &  $(1s 2p_{1/2})_1$  &      0.1587   &      0.0063   &      0.1650(2)   &         0.0260   &      0.0260     \\ 
       &  $(1s 2p_{3/2})_1$  &      0.1706   &      0.0062   &      0.1768(2)   &         0.0379   &      0.0378     \\ 
       &  $(1s 2p_{3/2})_2$  &      0.1452   &      0.0062   &      0.1515(2)   &         0.0125   &      0.0125     \\ 
       &   {\rm off-diag.}   &      0.0178   &      0.0000   &         0.0178   &         0.0178   &      0.0178     \\ 

\hline
   50  &           $(1s)^2$  &      0.3067   &      0.0286   &      0.3354(6)   &      0.1659(3)   &      0.1659     \\ 
       &        $(1s 2s)_0$  &      0.1939   &      0.0168   &      0.2106(3)   &      0.0412(1)   &      0.0412     \\ 
       &        $(1s 2s)_1$  &      0.1942   &      0.0168   &      0.2110(3)   &      0.0415(1)   &      0.0415     \\ 
       &  $(1s 2p_{1/2})_0$  &      0.1721   &      0.0144   &      0.1865(3)   &      0.0170(1)   &      0.0170     \\ 
       &  $(1s 2p_{1/2})_1$  &      0.1866   &      0.0144   &      0.2010(3)   &         0.0316   &      0.0315     \\ 
       &  $(1s 2p_{3/2})_1$  &      0.1999   &      0.0143   &      0.2142(3)   &         0.0447   &      0.0447     \\ 
       &  $(1s 2p_{3/2})_2$  &      0.1704   &      0.0143   &      0.1847(3)   &         0.0152   &      0.0152     \\ 
       &   {\rm off-diag.}   &      0.0207   &      0.0000   &         0.0207   &         0.0207   &      0.0206     \\ 

\hline
   60  &           $(1s)^2$  &      0.3721   &      0.0622   &     0.4342(11)   &      0.2152(6)   &      0.2152     \\ 
       &        $(1s 2s)_0$  &      0.2360   &      0.0367   &      0.2726(6)   &      0.0536(2)   &      0.0537     \\ 
       &        $(1s 2s)_1$  &      0.2364   &      0.0367   &      0.2731(6)   &      0.0540(1)   &      0.0541     \\ 
       &  $(1s 2p_{1/2})_0$  &      0.2103   &      0.0316   &      0.2418(6)   &      0.0228(2)   &      0.0227     \\ 
       &  $(1s 2p_{1/2})_1$  &      0.2276   &      0.0316   &      0.2591(5)   &      0.0400(1)   &      0.0400     \\ 
       &  $(1s 2p_{3/2})_1$  &      0.2426   &      0.0312   &      0.2737(5)   &         0.0547   &      0.0546     \\ 
       &  $(1s 2p_{3/2})_2$  &      0.2071   &      0.0312   &      0.2383(5)   &         0.0192   &      0.0193     \\ 
       &   {\rm off-diag.}   &      0.0247   &      0.0000   &         0.0247   &         0.0247   &      0.0246     \\ 

\hline
   70  &           $(1s)^2$  &      0.4094   &      0.1176   &     0.5270(18)   &     0.2615(10)   &      0.2612     \\ 
       &        $(1s 2s)_0$  &      0.2610   &      0.0698   &     0.3308(10)   &      0.0653(2)   &      0.0658     \\ 
       &        $(1s 2s)_1$  &      0.2615   &      0.0698   &     0.3313(10)   &      0.0658(2)   &      0.0662     \\ 
       &  $(1s 2p_{1/2})_0$  &      0.2338   &      0.0601   &      0.2939(9)   &      0.0284(3)   &      0.0283     \\ 
       &  $(1s 2p_{1/2})_1$  &      0.2523   &      0.0601   &      0.3125(9)   &      0.0470(1)   &      0.0469     \\ 
       &  $(1s 2p_{3/2})_1$  &      0.2674   &      0.0590   &      0.3264(9)   &         0.0609   &      0.0608     \\ 
       &  $(1s 2p_{3/2})_2$  &      0.2287   &      0.0590   &      0.2878(9)   &         0.0223   &      0.0225     \\ 
       &   {\rm off-diag.}   &      0.0266   &      0.0000   &      0.0266(1)   &      0.0266(1)   &      0.0266     \\ 

\hline
   80  &           $(1s)^2$  &      0.4513   &      0.2209   &     0.6721(32)   &     0.3340(17)   &      0.3326     \\ 
       &        $(1s 2s)_0$  &      0.2897   &      0.1322   &     0.4219(19)   &      0.0838(4)   &      0.0853     \\ 
       &        $(1s 2s)_1$  &      0.2903   &      0.1322   &     0.4225(18)   &      0.0844(3)   &      0.0858     \\ 
       &  $(1s 2p_{1/2})_0$  &      0.2613   &      0.1139   &     0.3752(17)   &      0.0371(5)   &      0.0369     \\ 
       &  $(1s 2p_{1/2})_1$  &      0.2813   &      0.1139   &     0.3952(16)   &      0.0571(2)   &      0.0570     \\ 
       &  $(1s 2p_{3/2})_1$  &      0.2959   &      0.1110   &     0.4069(16)   &         0.0688   &      0.0686     \\ 
       &  $(1s 2p_{3/2})_2$  &      0.2535   &      0.1110   &     0.3645(16)   &         0.0264   &      0.0267     \\ 
       &   {\rm off-diag.}   &      0.0289   &      0.0000   &      0.0289(1)   &      0.0289(1)   &      0.0289     \\ 

\hline
   90  &           $(1s)^2$  &      0.4739   &      0.4071   &     0.8810(87)   &     0.4384(44)   &      0.4338     \\ 
       &        $(1s 2s)_0$  &      0.3074   &      0.2462   &     0.5536(50)   &      0.1110(8)   &      0.1168     \\ 
       &        $(1s 2s)_1$  &      0.3081   &      0.2462   &     0.5542(50)   &      0.1117(7)   &      0.1174     \\ 
       &  $(1s 2p_{1/2})_0$  &      0.2802   &      0.2121   &     0.4923(45)   &     0.0497(10)   &      0.0496     \\ 
       &  $(1s 2p_{1/2})_1$  &      0.3009   &      0.2121   &     0.5130(44)   &      0.0705(4)   &      0.0707     \\ 
       &  $(1s 2p_{3/2})_1$  &      0.3140   &      0.2048   &     0.5188(43)   &      0.0762(1)   &      0.0759     \\ 
       &  $(1s 2p_{3/2})_2$  &      0.2686   &      0.2048   &     0.4734(43)   &      0.0308(1)   &      0.0314     \\ 
       &   {\rm off-diag.}   &      0.0302   &      0.0000   &      0.0302(2)   &      0.0302(2)   &      0.0305     \\ 

\hline
   92  &           $(1s)^2$  &      0.4752   &      0.4604   &      0.936(11)   &     0.4657(58)   &      0.4600     \\ 
       &        $(1s 2s)_0$  &      0.3091   &      0.2790   &     0.5881(66)   &     0.1183(10)   &      0.1260     \\ 
       &        $(1s 2s)_1$  &      0.3098   &      0.2790   &     0.5888(66)   &      0.1190(9)   &      0.1266     \\ 
       &  $(1s 2p_{1/2})_0$  &      0.2825   &      0.2404   &     0.5229(59)   &     0.0531(11)   &      0.0531     \\ 
       &  $(1s 2p_{1/2})_1$  &      0.3033   &      0.2404   &     0.5438(58)   &      0.0739(4)   &      0.0743     \\ 
       &  $(1s 2p_{3/2})_1$  &      0.3159   &      0.2316   &     0.5476(57)   &      0.0777(1)   &      0.0774     \\ 
       &  $(1s 2p_{3/2})_2$  &      0.2701   &      0.2316   &     0.5017(57)   &      0.0318(1)   &      0.0324     \\ 
       &   {\rm off-diag.}   &      0.0304   &      0.0000   &      0.0304(2)   &      0.0304(2)   &      0.0308     \\ 

\hline
  100  &           $(1s)^2$  &      0.4793   &      0.7892   &      1.268(43)   &      0.632(22)   &      0.6180     \\ 
       &        $(1s 2s)_0$  &      0.3162   &      0.4832   &      0.799(25)   &     0.1633(33)   &      0.1895     \\ 
       &        $(1s 2s)_1$  &      0.3170   &      0.4832   &      0.800(25)   &     0.1640(32)   &      0.1902     \\ 
       &  $(1s 2p_{1/2})_0$  &      0.2933   &      0.4166   &      0.710(22)   &     0.0737(17)   &      0.0759     \\ 
       &  $(1s 2p_{1/2})_1$  &      0.3151   &      0.4166   &      0.732(22)   &      0.0955(6)   &      0.0984     \\ 
       &  $(1s 2p_{3/2})_1$  &      0.3261   &      0.3970   &      0.723(22)   &      0.0869(1)   &      0.0866     \\ 
       &  $(1s 2p_{3/2})_2$  &      0.2766   &      0.3970   &      0.674(22)   &      0.0375(1)   &      0.0382     \\ 
       &   {\rm off-diag.}   &      0.0320   &      0.0000   &      0.0320(3)   &      0.0320(3)   &      0.0328     \\ 

\hline

\end{longtable}


%% file: recoil_he_rom_v6_sub.bbl
\begin{thebibliography}{10}

\bibitem{Stoehlker:2000:3109}
{\relax Th}.{~}St\"ohlker, P.~H.{~}Mokler, F.{~}Bosch, R.~W.{~}Dunford,
  F.{~}Franzke, O.{~}Klepper, C.{~}Kozhuharov, T.{~}Ludziejewski, F.{~}Nolden,
  H.{~}Reich, P.{~}Rymuza, Z.{~}Stachura, M.{~}Steck, P.{~}Swiat, and
  A.{~}Warczak,
\newblock Phys. Rev. Lett. {\bf 85},~3109 (2000).

\bibitem{Gumberidze:2005:223001}
A.{~}Gumberidze, {\relax Th}.{~}St\"ohlker, D.{~}Bana\ifmmode~\acute{s}\else
  \'{s}\fi{}, K.{~}Beckert, P.{~}Beller, H.~F.{~}Beyer, F.{~}Bosch,
  S.{~}Hagmann, C.{~}Kozhuharov, D.{~}Liesen, F.{~}Nolden, X.{~}Ma,
  P.~H.{~}Mokler, M.{~}Steck, D.{~}Sierpowski, and S.{~}Tashenov,
\newblock Phys. Rev. Lett. {\bf 94},~223001 (2005).

\bibitem{Schweppe:1991:1434}
J.{~}Schweppe, A.{~}Belkacem, L.{~}Blumenfeld, N.{~}Claytor, B.{~}Feinberg,
  H.{~}Gould, V.~E.{~}Kostroun, L.{~}Levy, S.{~}Misawa, J.~R.{~}Mowat, and M.~H.{~}Prior,
\newblock Phys. Rev. Lett. {\bf 66},~1434 (1991).

\bibitem{Brandau:2003:073202}
C.{~}Brandau, C.{~}Kozhuharov, A.{~}M\"uller, W.{~}Shi, S.{~}Schippers,
  T.{~}Bartsch, S.{~}B\"ohm, C.{~}B\"ohme, A.{~}Hoffknecht, H.{~}Knopp,
  N.{~}Gr\"un, W.{~}Scheid, T.{~}Steih, F.{~}Bosch, B.{~}Franzke,
  P.~H.{~}Mokler, F.{~}Nolden, M.{~}Steck, {\relax Th}.{~}St\"ohlker, and
  Z.{~}Stachura,
\newblock Phys. Rev. Lett. {\bf 91},~073202 (2003).

\bibitem{Beiersdorfer:2005:233003}
P.{~}Beiersdorfer, H.{~}Chen, D.~B.{~}Thorn, and E.{~}Tr\"abert,
\newblock Phys. Rev. Lett. {\bf 95},~233003 (2005).

\bibitem{Yerokhin:2006:253004}
V.~A.{~}Yerokhin, P.{~}Indelicato, and V.~M.{~}Shabaev,
\newblock Phys. Rev. Lett. {\bf 97},~253004 (2006).

\bibitem{Kozhedub:2008:032501}
Y.~S.{~}Kozhedub, O.~V.{~}Andreev, V.~M.{~}Shabaev, I.~I.{~}Tupitsyn,
  C.{~}Brandau, C.{~}Kozhuharov, G.{~}Plunien, and {\relax Th}.{~}St\"ohlker,
\newblock Phys. Rev. A {\bf 77},~032501 (2008).

\bibitem{Chantler:2012:153001}
C.~T.{~}Chantler, M.~N.{~}Kinnane, J.~D.{~}Gillaspy, L.~T.{~}Hudson,
  A.~T.{~}Payne, L.~F.{~}Smale, A.{~}Henins, J.~M.{~}Pomeroy, J.~N.{~}Tan,
  J.~A.{~}Kimpton, E.{~}Takacs, and K.{~}Makonyi,
\newblock Phys. Rev. Lett. {\bf 109},~153001 (2012).

\bibitem{Artemyev:2005:062104}
A.~N.{~}Artemyev, V.~M.{~}Shabaev, V.~A.{~}Yerokhin, G.{~}Plunien, and G.{~}Soff,
\newblock Phys. Rev. A {\bf 71},~062104 (2005).

\bibitem{Drake:1988:586}
G.~W.{~}Drake,
\newblock Can. J. Phys. {\bf 66},~586 (1988).

\bibitem{Johnson:1992:R2197}
W.~R.{~}Johnson and J.{~}Sapirstein,
\newblock Phys. Rev. A {\bf 46},~R2197 (1992).

\bibitem{Chen:1993:3692}
M.~H.{~}Chen, K.~T.{~}Cheng, and W.~R.{~}Johnson,
\newblock Phys. Rev. A {\bf 47},~3692 (1993).

\bibitem{Plante:1994:3519}
D.~R.{~}Plante, W.~R.{~}Johnson, and J.{~}Sapirstein,
\newblock Phys. Rev. A {\bf 49},~3519 (1994).

\bibitem{Indelicato:1995:1132}
P.{~}Indelicato,
\newblock Phys. Rev. A {\bf 51},~1132 (1995).

\bibitem{Trassinelli:2007:129}
M.{~}Trassinelli, S.{~}Boucard, D.~S.{~}Covita, D.{~}Gotta, A.{~}Hirtl,
  P.{~}Indelicato, {\'E}.-O.{~}Le~Bigot, J.~M.~F.{~}dos~Santos, L.~M.{~}Simons,
  L.{~}Stingelin, J.~F.~C.~A.{~}Veloso, A.{~}Wasser, and J.{~}Zmeskal,
\newblock J.~Phys. Conf. Ser. {\bf 58},~129 (2007).

\bibitem{Trassinelli:2009:63001}
M.{~}Trassinelli, A.{~}Kumar, H.~F.{~}Beyer, P.{~}Indelicato, R.{~}M\"artin,
  R.{~}Reuschl, Y.~S.{~}Kozhedub, C.{~}Brandau, H.{~}Br\"auning, S.{~}Geyer,
  A.{~}Gumberidze, S.{~}Hess, P.{~}Jagodzinski, C.{~}Kozhuharov, D.{~}Liesen,
  U.{~}Spillmann, S.{~}Trotsenko, G.{~}Weber, D.~F.~A.{~}Winters, and T.{~}St\"ohlker,
\newblock Eur. Phys. Lett. {\bf 87},~63001 (2009).

\bibitem{Amaro:2012:043005}
P.{~}Amaro, S.{~}Schlesser, M.{~}Guerra, {\'E}.-O.{~}Le~Bigot, J.-M.{~}Isac,
  P.{~}Travers, J.~P.{~}Santos, C.~I.{~}Szabo, A.{~}Gumberidze, and P.{~}Indelicato,
\newblock Phys. Rev. Lett. {\bf 109},~043005 (2012).

\bibitem{Rudolph:2013:103002}
J.~K.{~}Rudolph, S.{~}Bernitt, S.~W.{~}Epp, R.{~}Steinbr\"ugge, C.{~}Beilmann,
  G.~V.{~}Brown, S.{~}Eberle, A.{~}Graf, Z.{~}Harman, N.{~}Hell,
  M.{~}Leutenegger, A.{~}M\"uller, K.{~}Schlage, H.-C.{~}Wille,
  H.{~}Yava\ifmmode~\mbox{\c{s}}\else \c{s}\fi{}, J.{~}Ullrich, and
  J.~R.{~}Crespo L\'opez-Urrutia,
\newblock Phys. Rev. Lett. {\bf 111},~103002 (2013).

\bibitem{Kubicek:2014:032508}
K.{~}Kubi\ifmmode~\check{c}\else \v{c}\fi{}ek, P.~H.{~}Mokler, V.{~}M\"ackel,
  J.{~}Ullrich, and J.~R.{~}Crespo L\'opez-Urrutia,
\newblock Phys. Rev. A {\bf 90},~032508 (2014).

\bibitem{Beiersdorfer:2015:032514}
P.{~}Beiersdorfer and G.~V.{~}Brown,
\newblock Phys. Rev. A {\bf 91},~032514 (2015).

\bibitem{Epp:2015:020502_R}
S.~W.{~}Epp, R.{~}Steinbr\"ugge, S.{~}Bernitt, J.~K.{~}Rudolph, C.{~}Beilmann,
  H.{~}Bekker, A.{~}M\"uller, O.~O.{~}Versolato, H.-C.{~}Wille,
  H.{~}Yava\ifmmode~\mbox{\c{s}}\else \c{s}\fi{}, J.{~}Ullrich, and
  J.~R.{~}Crespo L\'opez-Urrutia,
\newblock Phys. Rev. A {\bf 92},~020502(R) (2015).

\bibitem{Holmberg:2015:012509}
J.{~}Holmberg, S.{~}Salomonson, and I.{~}Lindgren,
\newblock Phys. Rev. A {\bf 92},~012509 (2015).

\bibitem{Machado:HeBe:preprint}
J.{~}Machado, C.~I.{~}Szabo, J.~P.{~}Santos, P.{~}Amaro, M.{~}Guerra,
  A.{~}Gumberidze, {Guojie Bian}, J.~M.{~}Isac, and P.{~}Indelicato,
\newblock https://hal.archives-ouvertes.fr/hal-01556577.

\bibitem{Shabaev:1985:394:note}
V.~M.{~}Shabaev,
\newblock Teor. Mat. Fiz. {\bf 63},~394 (1985),
\newblock [Theor. Math. Phys. {\bf 63}, 588 (1985)].

\bibitem{Shabaev:1988:107:note}
V.~M.{~}Shabaev,
\newblock Yad. Fiz. {\bf 47},~107 (1988),
\newblock [Sov. J. Nucl. Phys. {\bf 47}, 69 (1988)].

\bibitem{Pachucki:1995:1854}
K.{~}Pachucki and H.{~}Grotch,
\newblock Phys. Rev. A {\bf 51},~1854 (1995).

\bibitem{Shabaev:1998:59}
V.~M.{~}Shabaev,
\newblock Phys. Rev. A {\bf 57},~59 (1998).

\bibitem{Adkins:2007:042508}
G.~S.{~}Adkins, S.{~}Morrison, and J.{~}Sapirstein,
\newblock Phys. Rev. A {\bf 76},~042508 (2007).

\bibitem{TTGF}
V.~M.{~}Shabaev,
\newblock Phys. Rep. {\bf 356},~119 (2002).

\bibitem{Palmer:1987:5987}
C.~W.~P.{~}Palmer,
\newblock J. Phys. B: At. Mol. Phys. {\bf 20},~5987 (1987).

\bibitem{Shabaev:1994:1307}
V.~M.{~}Shabaev and A.~N.{~}Artemyev,
\newblock J. Phys. B: At. Mol. Opt. Phys. {\bf 27},~1307 (1994).

\bibitem{Tupitsyn:2003:022511}
I.~I.{~}Tupitsyn, V.~M.{~}Shabaev, J.~R.{~}Crespo L\'opez-Urrutia,
  I.{~}Dragani\ifmmode~\acute{c}\else \'{c}\fi{}, R.~S.{~}Orts, and J.{~}Ullrich,
\newblock Phys. Rev. A {\bf 68},~022511 (2003).

\bibitem{Orts:2006:103002}
R.~S.{~}Orts, Z.{~}Harman, J.~R.{~}Crespo L\'opez-Urrutia, A.~N.{~}Artemyev,
  H.{~}Bruhns, A.~J.~G.{~}Mart\'{\i}nez, U.~D.{~}Jentschura, C.~H.{~}Keitel,
  A.{~}Lapierre, V.{~}Mironov, V.~M.{~}Shabaev, H.{~}Tawara, I.~I.{~}Tupitsyn,
  J.{~}Ullrich, and A.~V.{~}Volotka,
\newblock Phys. Rev. Lett. {\bf 97},~103002 (2006).

\bibitem{Korol:2007:022103}
V.~A.{~}Korol and M.~G.{~}Kozlov,
\newblock Phys. Rev. A {\bf 76},~022103 (2007).

\bibitem{Gaidamauskas:2011:175003}
E.{~}Gaidamauskas, C.{~}Naz\'e, P.{~}Rynkun, G.{~}Gaigalas, P.{~}J\"onsson, and
  M.{~}Godefroid,
\newblock J. Phys. B: At. Mol. Opt. Phys. {\bf 44},~175003 (2011).

\bibitem{Naze:2013:2187}
C.{~}Naz\'e, E.{~}Gaidamauskas, G.{~}Gaigalas, M.{~}Godefroid, and P.{~}J\"onsson,
\newblock Comp. Phys. Comm. {\bf 184},~2187 (2013).

\bibitem{Zubova:2016:052502}
N.~A.{~}Zubova, A.~V.{~}Malyshev, I.~I.{~}Tupitsyn, V.~M.{~}Shabaev,
  Y.~S.{~}Kozhedub, G.{~}Plunien, C.{~}Brandau, and {\relax Th}.{~}St\"ohlker,
\newblock Phys. Rev. A {\bf 93},~052502 (2016).

\bibitem{Filippin:2017:042502}
L.{~}Filippin, J.{~}Biero\ifmmode~\acute{n}\else \'{n}\fi{}, G.{~}Gaigalas,
  M.{~}Godefroid, and P.{~}J\"onsson,
\newblock Phys. Rev. A {\bf 96},~042502 (2017).

\bibitem{Artemyev:1995:1884}
A.~N.{~}Artemyev, V.~M.{~}Shabaev, and V.~A.{~}Yerokhin,
\newblock Phys. Rev. A {\bf 52},~1884 (1995).

\bibitem{Artemyev:1995:5201}
A.~N.{~}Artemyev, V.~M.{~}Shabaev, and V.~A.{~}Yerokhin,
\newblock J. Phys. B: At. Mol. Opt. Phys. {\bf 28},~5201 (1995).

\bibitem{Shabaev:1998:4235}
V.~M.{~}Shabaev, A.~N.{~}Artemyev, T.{~}Beier, G.{~}Plunien, V.~A.{~}Yerokhin, and G.{~}Soff,
\newblock Phys. Rev. A {\bf 57},~4235 (1998).

\bibitem{Shabaev:1999:493}
V.~M.{~}Shabaev, A.~N.{~}Artemyev, T.{~}Beier, G.{~}Plunien, V.~A.{~}Yerokhin, and G.{~}Soff,
\newblock Phys. Scr. {\bf T80},~493 (1999).

\bibitem{Angeli:2013:69}
I.{~}Angeli and K.~P.{~}Marinova,
\newblock At. Data Nucl. Data Tables {\bf 99},~69 (2013).

\bibitem{Yerokhin:2015:033103}
V.~A.{~}Yerokhin and V.~M.{~}Shabaev,
\newblock J. Phys. Chem. Ref. Data {\bf 44},~033103 (2015).

\bibitem{splines:DKB}
V.~M.{~}Shabaev, I.~I.{~}Tupitsyn, V.~A.{~}Yerokhin, G.{~}Plunien, and G.{~}Soff,
\newblock Phys. Rev. Lett. {\bf 93},~130405 (2004).

\bibitem{Sapirstein:1996:5213}
J.{~}Sapirstein and W.~R.{~}Johnson,
\newblock J. Phys. B: At. Mol. Opt. Phys. {\bf 29},~5213 (1996).

\bibitem{Grotch:1969:350}
H.{~}Grotch and D.~R.{~}Yennie,
\newblock Rev. Mod. Phys. {\bf 41},~350 (1969).

\bibitem{Borie:1982:67}
E.{~}Borie and G.~A.{~}Rinker,
\newblock Rev. Mod. Phys. {\bf 54},~67 (1982).

\bibitem{Aleksandrov:2015:144004}
I.~A.{~}Aleksandrov, A.~A.{~}Shchepetnov, D.~A.{~}Glazov, and V.~M.{~}Shabaev,
\newblock J. Phys. B: At. Mol. Opt. Phys. {\bf 48},~144004 (2015).

\bibitem{Sanders:1969:84}
F.~C.{~}Sanders and C.~W.{~}Scherr,
\newblock Phys. Rev. {\bf 181},~84 (1969).

\bibitem{Aashamar:1970:3324}
K.{~}Aashamar, G.{~}Lyslo, and J.{~}Midtdal,
\newblock J. Chem. Phys. {\bf 52},~3324 (1970).

\bibitem{Wang:2012:1603}
M.{~}Wang, G.{~}Audi, A.~H.{~}Wapstra, F.~G.{~}Kondev, M.{~}MacCormick,
  X.{~}Xu, and B.{~}Pfeiffer,
\newblock Chin. Phys. C {\bf 36},~1603 (2012).

\end{thebibliography}
